\documentclass[10pt,showpacs,twocolumn,nofootinbib,aps,prd]{revtex4-2}
\usepackage{CJK}
\usepackage{amsfonts,amsmath,amssymb,mathrsfs}
\usepackage{mathptmx,mathtools,array,latexsym}
\usepackage[colorlinks=true,linkcolor=blue,citecolor=blue,urlcolor=blue]{hyperref}

\def\be{\begin{equation}}
\def\ee{\end{equation}}
\def\bea{\begin{eqnarray}}
\def\eea{\end{eqnarray}}
\def\nn{\nonumber}
\def\cA{\mathcal{A}}
\def\cQ{\mathcal{Q}}
\def\bV{\mathbb{V}}
\def\CP2{\mathbb{CP}^2}
\def\p{\partial}

\begin{document}
\begin{CJK*}{GBK}{song}

\title{Consistent mass formulas for higher even-dimensional Taub-NUT spacetimes and
their AdS counterparts}

\author{Di Wu}
\email{wdcwnu@163.com}

\author{Shuang-Qing Wu}
\email{Corresponding author: sqwu@cwnu.edu.cn}

\affiliation{School of Physics and Astronomy, China West Normal University, Nanchong,
Sichuan 637002, People's Republic of China}

\date{\today}

\begin{abstract}
Currently, there is a great deal of interest in the seeking of consistent thermodynamics of
the Lorentzian Taub-NUT spacetimes. Despite a lot of ``satisfactory'' efforts have been made,
all of these activities have been restricted to the four-dimensional cases, with the higher
even-dimensional cases remaining unexplored. The aim of this article is to fill the gap, for
the first time. To the end of this subject, we first adopt our own idea that ``The NUT charge
is a thermodynamical multi-hair" to investigate the consistent thermodynamics of $D = 6, 8,
10$ Lorentzian Taub-NUT spacetimes without a cosmological constant. Similarly to the $D = 4$
cases as did in our previous works, we find that the first law and Bekenstein-Smarr mass
formulas are perfectly satisfied if we still assign the secondary hair: $J_n = Mn$ as a
conserved charge in both mass formulas. Turning to the cases with a nonzero cosmological
constant, our treatment continues to work very well and all the results can be fairly
generalized to the corresponding Taub-NUT AdS spacetimes in higher even-dimensions, although
we do not know how to define and introduce a similar higher-dimensional version of the dual
(magnetic) mass that is well known in four dimensions. Based upon the preceding results, we
will also derive the reduced version of the mass formulas when the secondary hair $J_n$ is
viewed as a redundant thermodynamic variable.
\end{abstract}

\maketitle
\end{CJK*}

\section{Introduction}

Taub-NUT solutions \cite{AM53-472,JMP4-915} have long been a source of insight into
gravitational thermodynamics. The solutions possess a number of undesirable properties
that, while at first sight highly pathological, actually result in important clarifications
in our understanding of black hole thermodynamics, such as the geometrical interpretation
of entropy. Recently, there has been a resurgence of great interest in exploring the
consistent thermodynamics of the Lorentzian Taub-NUT spacetimes \cite{PRD100-101501,
PRD105-124013,2210.17504,PRD100-064055,JHEP0719119,CQG36-194001,PLB798-134972,JHEP0520084,
PRD100-104016,PLB832-137264,PLB802-135270,IJMPD31-2250021,JHEP0821152,PRD101-124011,
PRD105-124034,PRD103-024052,JHEP0321039,PRD106-024022,EPJP130-124,2112.00780,2208.05494}.
In our opinion, these current investigations of the first law of the NUT-charged spacetimes
can be categorized into three different schemes: (I) Retaining the mass unmodified and
introducing new global-like charges (secondary hairs) together with their conjugate
potentials \cite{PRD100-101501,PRD105-124013}; (II) Keeping the mass unchanged and
including new nonglobal Misner charges and their conjugate variables \cite{PRD100-064055,
PLB798-134972,JHEP0719119,JHEP0520084,CQG36-194001,PRD100-104016}; and (III) Only modifying
the mass by taking account for the contribution of new nonglobal charges \cite{PRD101-124011,
PRD105-124034}. Note that in Ref. \cite{2208.05494}, the thermodynamic mass that enters into
the first law of the four-dimensional Taub-NUT spacetime is the horizon mass \cite{PRD56-961}.
Besides these, there is fewer interest \cite{PRD100-064055,JHEP0821152} to consider the
entropy as the Noether charge \cite{CQG17-3317} that includes the horizon area and the
contribution from the Misner strings. However, all of the above-mentioned efforts are only
restricted to four-dimensional cases, leaving thermodynamics of the Lorentzian Taub-NUT
spacetimes in higher even-dimensions unexplored, which motivates the subject of the present
article.

In our previous papers \cite{PRD100-101501,PRD105-124013,2210.17504}, we have advocated a
new idea that ``The NUT charge is a thermodynamical multi-hair" and put forward a simple,
systematic way to study the consistent thermodynamics of almost all of the four-dimensional
(dyonic) NUT-charged spacetimes. It should be emphasized that, unlike all other attempts
\cite{PRD100-064055,JHEP0719119,CQG36-194001,PLB798-134972,JHEP0520084,PRD100-104016,
PLB832-137264,PLB802-135270,IJMPD31-2250021,JHEP0821152,PRD101-124011,PRD105-124034,
2112.00780,2208.05494}, our scheme only relies on deriving firstly a new meaningful
Christodoulou-Ruffini-type squared-mass formula \cite{PRL25-1596,PRD4-3552} satisfied
by the four-dimensional (dyonic) NUT-charged spacetimes, and the only needed input in
this derivation is to introduce the secondary hairs: ($J_n = Mn$, $Q_n = qn$ and $P_n = pn$)
as new conserved charges. Then the consistent thermodynamic first law and Bekenstein-Smarr
mass formulas of these NUT-charged spacetimes can be deduced via some simple and purely
algebraic manipulations from this squared-mass formula, which can hardly be given by the
other papers as mentioned above. Subsequently, the usual Bekenstein-Hawking one-quarter
area-entropy relation can be naturally restored for the generic NUT-charged spacetime
(and all its extensions) without imposing any constraint condition and with no need to
assume ahead that the one-quarter area-entropy relation should hold true. The advantage
of our proposal that the NUT charge acts as a thermodynamical multi-hair is that it can
not only explicate the rotation-like and electromagnetic charge-like characters, but
also simultaneously explain many other exotic properties. What is more, our consistent
mass formulas \cite{PRD100-101501,PRD105-124013} are unique, and all expressions for
thermodynamical quantities are exceedingly simple and succinct. This is in contrast to
all other works where not only can the consistent first law of the NUTty dyonic spacetimes
have the electric-type, magnetic-type, mixed-type versions \cite{JHEP0719119,JHEP0520084},
and even many other ones \cite{PRD100-104016}, but also the expressions of the related
thermodynamical variables are quite complicated.

In addition, it should be emphasized that the introduction of the secondary hair $J_n = Mn$
in our previous works \cite{PRD100-101501,PRD105-124013,2210.17504} is not merely based upon
thermodynamical reasons, but also come from many other considerations. For instance, our
secondary hair $J_n = Mn \equiv M_5$ must endow the character of a global conserved charge,
which exactly corresponds to the mass of the five-dimensional gravitational magnetic monopole
\cite{PLB634-531}, so that it can be naturally included into the first law and Bekenstein-Smarr
mass formula. On the other hand, it can not only help to explain the gyromagnetic ratio of a
Kerr-NUT-type spacetime \cite{PRD77-044038}, but also accounts for the quantization condition
for a gravitational monopole \cite{CQG3-65,PPS92-1,GRG5-603}. What's more, it has been later
shown in Ref. \cite{PLB807-135521} that only considering the secondary hair $J_n = Mn$ as
a independent charge, can the area (or entropy) products of the NUT-charged spacetimes be
subject to the universal rules \cite{PRL106-121301}, and the mass be expressed as a sum of
the surface energy, the rotational energy and the electromagnetic energy \cite{GRG53-69}.

In this work, we will continue to apply our proposal that ``The NUT charge is a thermodynamical
multi-hair" to investigate consistent thermodynamics of the $D = 6,8,10$ Lorentzian Taub-NUT
spacetimes without and with a cosmological constant. Our paper is organized as follows. In Sec.
\ref{s2}, we start with the construction of a novel Christodoulou-Ruffini-like squared-mass
formula of the six-dimensional Lorentzian Taub-NUT solution by additionally including only
one secondary hair $J_n = Mn$, as did in Refs. \cite{PRD100-101501,PRD105-124013}. Using this
squared-mass formula, which can be thought of as representing a hyper-surface embedded into
one more high dimensional thermodynamical state space, both the differential and integral
mass formulas can be deduced through a simple mathematical manipulation. Then, the procedure
is extended to the six-dimensional Lorentzian Taub-NUT-AdS case. In Sec. \ref{s3}, we proceed
to discuss the cases of the eight-dimensional Lorentzian Taub-NUT and Taub-NUT-AdS spacetimes,
respectively. Then, in Sec. \ref{s4}, we extend to investigate the cases of the ten-dimensional
Taub-NUT spacetime and its AdS extension. We find that our scheme in the $D = 6,8,10$ cases
works successfully as in the four-dimensional case \cite{PRD100-101501}, and summarize in
Sec. \ref{s5} the main results for the generic $(2k+2)$-dimensional Taub-NUT-AdS spacetimes.
In Sec. \ref{s6}, we will turn to consider the secondary hair $J_n$ as a redundant  thermodynamic
variable and derive the corresponding mass formulas for all ($2k+2$)-dimensions when $J_n$
is not considered as a independent thermodynamic variable. Finally, we present our conclusions
and outlooks in Sec. \ref{s7}. In the Appendix \ref{app}, we briefly present the main results of
our extensions to the cases of the ($2k+2$)-dimensional multi-NUTty spacetimes without a
cosmological constant.

\section{6-dimensional Taub-NUT spacetime}\label{s2}

As shown in Ref. \cite{CQG19-2051} for the six-dimensional Taub-NUT spacetime, there are two
different choices for the base space, namely, $S^2 \times{} S^2$ and $\CP2$. We start our
investigation of the mass formulas in the case of the $S^2 \times S^2$ base space, but the
same procedure is also applicable to the case of the $\CP2$ base space. Using $S^2\times{}
S^2$ as a base space, the metric of the six-dimensional Lorentzian Taub-NUT solution has the
form:
\bea
ds_6^2 &=& -f(r)\Big(dt +2n\sum_{i=1}^2\cos\theta_i{}d\phi_i\Big)^2
 +\frac{dr^2}{f(r)} \nn \\
&&+\big(r^2 +n^2\big)\sum_{i=1}^2\big(d\theta_i^2
 +\sin^2\theta_i{}d\phi_i^2\big) \, , \quad \label{6dNUT}
\eea
where

\be
f(r) = \frac{r^4 +6n^2r^2 -3n^4 -6mr}{3\big(r^2 +n^2\big)^2} \, , \nn
\ee
in which $m$ and $n$ are the mass parameter and the NUT charge parameter, respectively.

Our aim is to derive various mass formulas and to discuss consistent thermodynamics of the
six-dimensional Lorentzian Taub-NUT spacetime. To begin with, let us present some known
quantities that can be evaluated via the standard method. First, the area and the surface
gravity at the horizon are easily computed as
\be
A_h = 16\pi^2\big(r_h^2 +n^2\big)^2 = 16\pi^2\cA_h \, , ~
\kappa = \frac{1}{2}f^{\prime}(r_h) = \frac{1}{2r_h} \, , \quad \label{6dAk}
\ee
in which a reduced horizon area $\cA_h  = (r_h^2 +n^2)^2 $ is introduced just for briefness,
and $r_h$ represents the greatest root of the horizon equation: $r_h^4 +6n^2r_h^2 -3n^4
-6mr_h = 0$.

As for the global conserved charges ($M$ and $N$), the Komar mass is divergent, while the
Abbott-Deser (AD) mass \cite{NPB195-76} is finite. The AD mass $M$ associated to the
Killing vector $\p_t$ and the NUT charge $N$ read
\be
M = 8\pi{}m \, , \qquad N = 8\pi{}n \, . \label{MN6}
\ee

In addition to the above global conserved charges $(M, N)$ which act as the primary hairs,
below just as did in the four-dimensional cases \cite{PRD100-101501,PRD105-124013,2210.17504},
we will also simply introduce an extra secondary hair $J_n \simeq mn$ into the mass formula,
which appears in the following asymptotic expansions of the metric components $g_{t\phi_1}$
and $g_{t\phi_2}$ at infinity:
\bea
&&g_{tt} \simeq -\frac{1}{3} -\frac{4n^2}{3r^2} +\frac{2m}{r^3}
 +\mathcal{O}(r^{-4}) \, , \nn \\
&&g_{t\phi_i} \simeq 2n{}g_{tt}\cos\theta_i \, , \qquad i=1, 2 \, .
\eea
It should be mentioned that the hairs ($M, N, J_n$) appeared in the first law and the
Bekenstein-Smarr mass formula are the lowest three moments of the multi-pole moments:
$m^in^j$ or $(m +In)^k$ where $i\geq 0, j\geq 0, k\geq 0$ are non-negative integers. In
particular, $J_n$ is the only introduced secondary hair so that all the thermodynamical
expressions of the solutions will not be rather complicated in our work.

\subsection{Consistent mass formulas of the 6-dimensional Taub-NUT spacetime}

In order to establish the first law which is reasonable and consistent in both physical and
mathematical senses, we employ the algebraic approach suggested in Refs. \cite{PRD100-101501,
PRD105-124013,PLB608-251} to construct a meaningful Christodoulou-Ruffini-type squared-mass
formula. First, via reexpressing $r_h = \sqrt{\cA_h^{1/2} -n^2}$ in terms of the reduced
horizon area and substituting it into the equation: $(r_h^4 +6n^2r_h^2 -3n^4)^2 = 36m^2r_h^2$,
we get the following identity:

\be
m^2 = \frac{1}{36\sqrt{\cA_h}}\big(\cA_h +4n^2\sqrt{\cA_h} -8n^4\big)^2
 +\frac{m^2n^2}{\sqrt{\cA_h}} \, , \label{sqm}
\ee
which can be alternatively converted to a quartic polynomial of $\cA_h$:
\be
\big(\cA_h^2 +36m^2n^2 +64n^8\big)^2
= 16\big(9m^2 +16n^6 -2n^2\cA_h \big)^2\cA_h \, . \nn \quad
\ee

Next, in addition to the conserved charges $M$ and $N$ given in Eq. (\ref{MN6}), only one
extra input that we need is to introduce the secondary hair $J_n = Mn = 8\pi{}mn$ as a
thermodynamic independent variable. Then after substituting $m = M/(8\pi)$, $n = N/(8\pi)$
and $\cA = 8\pi\cA_h$ into Eq. (\ref{sqm}), one can arrive at an useful identity
\be
M^2 = \frac{\sqrt{2\pi}}{18\sqrt{\cA}}\Big(\cA +\frac{N^2}{8\pi^2}\sqrt{2\pi\cA}
 -\frac{N^4}{64\pi^3}\Big)^2 +\frac{2\sqrt{2\pi}}{\sqrt{\cA}}J_n^2\, , \label{sqM}
\ee
which is our new Christodoulou-Ruffini-like squared-mass formula for the six-dimensional
Taub-NUT spacetime. Alternatively, the above equation (\ref{sqM}) can be converted to a
quartic polynomial of the area $\cA = \cA(M, N, J_n)$:
\bea
\Big(\cA^2 +36J_n^2 +\frac{N^8}{4096\pi^6}\Big)^2
 = \frac{\cA}{8\pi^3}\Big(N^2\cA -36\pi{}M^2 -\frac{N^6}{64\pi^3} \Big)^2 \, . \nn
\eea

At this step, it would be stressed that Eq. (\ref{sqm}) can be thought of as representing
a hyper-surface in the three-dimensional thermodynamical state space, whose variables ($m,
n, \cA_h$) exactly match with the numbers of the solution parameters that appeared in the
structure function $f(r)$. After introducing an extra hair $J_n$, which is nothing but a
kind of higher-dimensional embedding trick, it becomes a hyper-surface in the four-dimensional
state space, as specified by Eq. (\ref{sqM}), which now has four independent variables ($M,
N, J_n, \cA$). Our below discussions will be based upon this higher one-dimensional
thermodynamical state space.

Having finished the above task, we are now in a position to obtain the differential and integral
mass formulas for the six-dimensional Taub-NUT spacetime. Since the secondary hair $J_n$ will
be treated as an independent variable,\footnote{However, one may object to this viewpoint. A
treatment without viewing it as a independent thermodynamic variable in the mass formulas for
all ($2k+2$)-dimensions is presented in Sec. \ref{s6}.} the above squared-mass formula (\ref{sqM})
can be regarded formally as a basic functional relation: $M = M(\cA, N, J_n)$. As did in Refs.
\cite{PRD100-101501,PRD105-124013,2210.17504,PRD103-044014,PRD101-024057,PRD102-044007},
differentiating it with respect to the thermodynamical variables ($\cA, N, J_n$) yields their
conjugate quantities, and subsequently we can arrive at the differential and integral mass
formulas with the conjugate thermodynamic potentials given by the ordinary Maxwell relations.

For instance, differentiating the squared-mass formula (\ref{sqM}) with respect to $\cA$
yields one-quarter of the surface gravity:
\be
\kappa = 4\frac{\p{}M}{\p\cA}\Big|_{(N,J_n)} = \frac{1}{2r_h} \, ,
\ee
which is exactly the same one as given in Eq. (\ref{6dAk}). Similarly, by differentiating
the squared-mass formula (\ref{sqM}) with respect to the NUT charge $N$ and the secondary
hair $J_n$, then their conjugate gravito-magnetic potential $\psi_h$ and quasi-angular
momentum $\omega_h$ can be derived, respectively, as follows:
\bea
&&\psi_h = \frac{\p{}M}{\p{}N}\Big|_{(\cA,J_n)}
 = \frac{4nr_h\big(r_h^2 -3n^2\big)}{3\big(r_h^2 +n^2\big)} \, , \\
&&\omega_h = \frac{\p{}M}{\p{}J_n}\Big|_{(\cA,N)} = \frac{n}{r_h^2 +n^2} \, .
\eea

Now, one can check that both the differential and integral mass formulas are completely
fulfilled
\bea
dM &=& (\kappa/4)d\cA +\omega_h{}dJ_n +\psi_h{}dN \, ,  \label{dmf} \\
3M &=& \kappa\cA +4\omega_h{}J_n +\psi_h{}N \, , \label{imf}
\eea
among all the aforementioned thermodynamical conjugate pairs. Comparing these mass formulas
(\ref{dmf}-\ref{imf}) with the standard ones, it is highly urged that the following familiar
identifications be made:
\be
S = \frac{A_h}{4} = \frac{\pi}{2}\cA = 4\pi^2\big(r_h^2 +n^2\big)^2 \, , \qquad
T = \frac{\kappa}{2\pi} = \frac{1}{4\pi{}r_h} \, , \label{ts}
\ee
which naturally recovers the famous Bekenstein-Hawking one-quarter area-entropy relation
of the six-dimensional Taub-NUT spacetime, completely similar to the $D = 4$ cases.

\subsection{Extension to the Taub-NUT-AdS$_6$ spacetime}\label{IIB}

Now we will extend the above discussion to explore the Lorentzian Taub-NUT-AdS$_6$ spacetime
with a nonzero negative cosmological constant. The metric is still given by Eq. (\ref{6dNUT}),
but now we have
\bea
f(r) &=& \frac{1}{3\big(r^2 +n^2\big)^2}\Big[r^4 +6n^2r^2 -3n^4 -6mr \nn \\
&&+3g^2\big(r^6 +5n^2r^4 +15n^4r^2 -5n^6\big) \Big] \, , \nn
\eea
where $l = 1/g$ is the cosmological scale.

First, we will employ the conformal completion method \cite{PRD73-104036} to calculate the
conserved mass $M$ of the Taub-NUT-AdS$_6$ solution. This conformal AMD mass can be evaluated
via the integral in terms of the conformal Weyl tensor over the spatial conformal boundary
at infinity. The Taub-NUT-AdS$_6$ spacetime is asymptotically local AdS, and admits an
asymptotic boundary 5-metric that approaches to
\bea
d\bar{s}_5^2 = \lim_{r\to\infty}\frac{ds_6^2}{r^2} &=& -g^2\Big(dt
 +2n\sum_{i=1}^2\cos\theta_i{}d\phi_i\Big)^2 \nn \\
&&+\sum_{i=1}^2\big(d\theta_i^2 +\sin^2\theta_i{}d\phi_i^2\big) \, , \quad  \label{bm}
\eea
with which one can define a normal vector: $\hat{n}^a = -gr^2\delta_r^a$.

Note that the 5-volume form of the conformal boundary AdS metric (\ref{bm}) is simply given by
\newpage
\be
\bV_5 = g\sin\theta_1\sin\theta_2\, dt \wedge{} d\theta_1 \wedge{}
d\theta_2 \wedge{} d\phi_1 \wedge{} d\phi_2 \, ,
\ee
then using the inner-product rule $<\p_{\mu}, dx^{\mu}> = \delta_{\mu}^{\nu}$, we can obtain
the area vector: $d\Sigma_t = <\p_t, \bV_5> = g\sin\theta_1\sin\theta_2\,  d\theta_1 \wedge{}
d\theta_2 \wedge{} d\phi_1 \wedge{} d\phi_2$, from which we can get its only non-vanishing
component:
\be
dS_t = g\prod_{i=1}^2\sin\theta_i{}d\theta_i{}d\phi_i \, .
\ee

Since the conserved charge associated with a unit Killing vector $\xi^{\nu}$ is defined as
\be \label{AMDm}
\cQ[\xi] = \frac{1}{24\pi{}g^3}\int \big(r^3C^t_{~a{\nu}b}\hat{n}^a\hat{n}^b\xi^{\nu}
 dS_t\big)\big|_{r\to\infty} \, ,
\ee
where $C^t_{~a{\nu}b}$ is the Weyl conformal tensor, we can easily obtain the conserved
charge associated with the timelike vector $\p_t$ as:
\be
\cQ[\p_t] = 8\pi{}m -\frac{8\pi}{3}(1+6g^2n^2)n^2r +\mathcal{O}(r^{-1}) \, ,
\ee
which is clearly divergent at spatial infinity. Therefore, in order to obtain a finite
expression for the conformal mass $M = 8\pi{}m$, one must subtract the divergence due to
the contribution from the massless (pure NUT) background spacetime. So we see that while
the conformal completion method can get a finite expression for the NUT-AdS$_4$ spacetime,
it fails to do so in the higher even-dimensional NUT-charged AdS spacetimes \cite{JHEP0306083}.
This situation is very much similar to the Komar integral which can achieve a finite value
for the four-dimensional RN-NUT spacetime, while can not obtain a finite one for higher
even-dimensional NUT-charged spacetimes.

On the other hand, the Abbott-Deser method \cite{NPB195-76} is a reference background
subtraction approach, which fairly gives a finite AD mass \cite{PRD73-064020}. Incidentally,
one can also use the counterterm method \cite{NPB652-348,NPB674-329,JHEP0105049,PLB620-1}
to obtain the same result for the mass, but also to get a finite expression for the Euclidean
action at the same time. However we shall not adopt this method here due to its involved
computations.

Unfortunately, since it is unclear to us how to define a dual (magnetic) mass in the higher
dimensional spacetime, we will not consider the dual mass here and hereafter. The NUT charge
will be simply taken as $N = 8\pi{}n$ just like the case without a cosmological constant.

Next, the surface gravity at the horizon which is specified by the largest root of equation:
$f(r_h) = 0$ can be evaluated as
\be
\kappa = \frac{1}{2}f^{\prime}(r_h) = \frac{1 +5g^2\big(r_h^2 +n^2\big)}{2r_h} \, ,
\label{6dkappa}
\ee
while the horizon area reads: $A_h = 16\pi^2\cA_h$, in which the reduced horizon area is
still denoted as: $\cA_h = (r_h^2 +n^2)^2$.

Now we would like to derive a novel Christodoulou-Ruffini-like squared-mass formula like the
case without a cosmological constant. Accordingly, inserting $r_h = \sqrt{\cA_h^{1/2} -n^2}$
into the equation: $[r_h^4 +6n^2r_h^2 -3n^4 +3g^2(r_h^6 +5n^2r_h^4 +15n^4r_h^2 -5n^6)]^2 =
36m^2r_h^2$ will yield

\bea
m^2 &=& \frac{1}{36\sqrt{\cA_h}}\Big[\big(1 +6g^2n^2\big)
 \big(\cA_h +4n^2\sqrt{\cA_h} -8n^4\big) \nn \\
&&+3g^2\cA_h^{3/2}\Big]^2 +\frac{m^2n^2}{\sqrt{\cA_h}} \, , \label{6dNUTAdSsqm}
\eea
which can be converted to a sextic polynomial of $\cA_h$:
\bea
&&\Big[9g^4\cA_h^3 +\big(1 +6g^2n^2\big)\big(1 +30g^2n^2\big)\cA_h^2
 +64n^8\big(1 +6g^2n^2\big)^2 \nn \\
&&+36m^2n^2\Big]^2 = 4\Big[18m^2 +\big(1 +6g^2n^2\big)\big(3g^2\cA_h +24g^2n^4 \nn \\
&&\qquad\qquad\qquad  +4n^2\big)\big(8n^4-\cA_h\big)\Big]^2\cA_h \, .
\eea

Finally, plugging $m = M/(8\pi)$, $n = N/(8\pi)$, $\cA = 8\pi\cA_h$ and $g^2 = 4\pi{}P/5$
into Eq. (\ref{6dNUTAdSsqm}), where $P = (D-1)(D-2)g^2/(16\pi)$ is the generalized pressure
\cite{PRD84-024037}, and also introducing a secondary hair: $J_n = Mn$ as before, then after
a little algebra we obtain an useful identity:
\bea
M^2 &=& \frac{\sqrt{2\pi}}{18\sqrt{\cA}}\bigg[\Big(1 +\frac{3N^2}{40\pi}P\Big)\Big(\cA
 +\frac{N^2}{8\pi^2}\sqrt{2\pi\cA} -\frac{N^4}{64\pi^3}\Big) \nn \\
&&+\frac{3}{10\pi}(2\pi\cA)^{3/2}P \bigg]^2
 +\frac{2\sqrt{2\pi}}{\sqrt{\cA}}J_n^2  \, , \label{SQM}
\eea
which is nothing but the Christodoulou-Ruffini-like squared-mass formula for the six-dimensional
Taub-NUT-AdS spacetime. Eq. (\ref{SQM}) consistently reduces to Eq. (\ref{sqM}) obtained in the
case of the six-dimensional Taub-NUT spacetime when the generalized pressure $P$ is turned off.

Like the case without a cosmological constant, Eq. (\ref{6dNUTAdSsqm}) represents a
hyper-surface in the four-dimensional state space with four free variables ($m, n, g, \cA_n$).
After introducing an extra hair $J_n$, it is embedded into a five-dimensional thermodynamic
state space defined by Eq. (\ref{SQM}), which is our starting point of the following
prescription.

The differentiation of the squared-mass formula (\ref{SQM}) leads to the first law:
\be
dM = (\kappa/4)d\cA +\omega_h{}dJ_n +\psi_h{}dN +VdP \, ,  \label{FL}
\ee
where
\bea
\kappa &=& 4\frac{\p{}M}{\p\cA}\Big|_{(N,J_n,P)}
 = \frac{1 +5g^2\big(r_h^2 +n^2\big)}{2r_h} \, , \nn \\
\omega_h &=& \frac{\p{}M}{\p{}J_n}\Big|_{(\cA,N,P)} = \frac{n}{r_h^2 +n^2} \, , \nn\\
\psi_h &=& \frac{\p{}M}{\p{}N}\Big|_{(\cA,J_n,P)} \nn \\
&=& \frac{2nr_h\big[2r_h^2 -6n^2 +3g^2\big(r_h^4 +10n^2r_h^2
 -15n^4\big)\big]}{3\big(r_h^2 +n^2\big)} \, , \nn \\
V &=& \frac{\p{}M}{\p{}P}\Big|_{(\cA,N,J_n)} =
 \frac{16\pi^2r_h(r_h^6 +5n^2r_h^4 +15n^4r_h^2 -5n^6)}{5\big(r_h^2 +n^2\big)} \, . \nn
\eea
When the NUT charge parameter $n$ vanishes, the thermodynamic volume reduces to
$V = 16\pi^2r_h^5/5$.

Utilizing all the expressions obtained above, one can directly verify that the Bekenstein-Smarr
mass formula
\be
3M = \kappa\cA +4\omega_h{}J_n +\psi_h{}N -2VP \, ,
\ee
is completely satisfied also. It is naturally suggested to identify $S = A_h/4 = 4\pi^2\cA_h$
and $T = \kappa/(2\pi)$, so that the solution acts like a genuine black hole without breaking
the classical one-quarter area-entropy relation.

In the remaining two sections, we will adopt the same strategy to deal with the eight- and
ten-dimensional NUT-charged (AdS) spacetimes, respectively. The interpretation of our
squared-mass formulas in both dimensions is essentially the same one as did just in the
six-dimensional case, and will not be repeated once more again.

\section{8-dimensional Taub-NUT spacetime}\label{s3}

In this section, we will extend the above discussion to the case of the eight-dimensional
Taub-NUT spacetime, to which there are two different choices \cite{CQG19-2051} for the base
manifold, namely $S^2 \times{} S^2 \times{} S^2$ and $S^2\times{}\CP2$. Likewise the
six-dimensional case, we will only consider the case where the base space is $S^2\times{} S^2
\times{} S^2$, so that the metric owns a $U(1)$ fibration over $S^2\times{} S^2\times{} S^2$:
\bea
ds_8^2 &=& -f(r)\Big(dt +2n\sum_{i=1}^3\cos\theta_i{}d\phi_i\Big)^2 +\frac{dr^2}{f(r)} \nn \\
&&+\big(r^2 +n^2\big)\sum_{i=1}^3\big(d\theta_i^2 +\sin^2\theta_i{}d\phi_i^2\big) \, ,
\label{8dNUT}
\eea
where
\be
f(r) = \frac{r^6 +5n^2r^4 +15n^4r^2 -5n^6 -10mr}{5\big(r^2 +n^2\big)^3} \, . \nn
\ee

At the horizon which is the largest root of $f(r_h) = 0$, the area and the surface gravity
can be evaluated via the standard method as
\be
A_h = 64\pi^3\big(r_h^2 +n^2\big)^3 = 64\pi^3\cA_h \, , \quad
\kappa = \frac{1}{2}f^{\prime}(r_h) = \frac{1}{2r_h} \, , \qquad  \label{8dAk}
\ee
where we now denote the reduced horizon area: $\cA_h = (r_h^2 +n^2)^3$.

Similar to the six-dimensional case, the AD mass and the NUT charge can be computed as:
\be
M = 48\pi^2m \, , \qquad N = 48\pi^2n \, .
\ee

\subsection{Consistent mass formulas of the 8-dimensional Taub-NUT spacetime}

To derive our squared mass formula, we will adopt the same trick as did in the last section,
so we first express the positive root $r_h = \sqrt{\cA_h^{1/3} -n^2}$ in terms of the reduced
horizon area and substitute it into the equation: $(r_h^6 +5n^2r_h^4 +15n^4r_h^2 -5n^6)^2 =
100m^2r_h^2$. After some algebraic computations, one can obtain the following useful identity:

\bea
m^2 &=& \frac{1}{100\cA_h^{1/3}}\big(\cA_h +2n^2\cA_h^{2/3} +8n^4\cA_h^{1/3}
 -16n^6\big)^2 \nn \\
&& +\frac{m^2n^2}{\cA_h^{1/3}} \, , \label{8dsqm}
\eea
which can also be converted into a polynomial of $\cA_h$ after eliminating the fractional
powers. Due to its complexity, we shall omit it here.

Subsequently, after inserting $m = M/(48\pi^2)$, $n = N/(48\pi^2)$ and $\cA = 48\pi^2\cA_h$
into Eq. (\ref{8dsqm}) and including only one secondary hair: $J_n =  Mn$ as before, we can
obtain a novel squared-mass formula:
\bea
M^2 &=& \frac{\big(6\pi^2\big)^{1/3}}{50\cA^{1/3}}\Big[\cA
 +\frac{N^2\big(6\pi^2\cA^2\big)^{1/3}}{576\pi^4}
 +\frac{N^4(36\pi\cA)^{1/3}}{165888\pi^7} \nn \\
&&-\frac{N^6}{15925248\pi^{10}}\Big]^2 +\frac{2\big(6\pi^2\big)^{1/3}}{\cA^{1/3}}J_n^2 \, .
\label{8dsqM}
\eea

Now we employ a similar procedure as manipulated in the previous section, i.e., viewing the
secondary hair $J_n = Mn$ as an independent thermodynamical variable, then performing the
partial derivative of the above squared-mass formula (\ref{8dsqM}) with respect to one of
its thermodynamical quantities ($\cA, N, J_n$) and simultaneously fixing the remaining ones,
respectively, and this will lead to their corresponding conjugate quantities.

First, differentiating the squared-mass formula (\ref{8dsqM}) with respect to $\cA$ yields
one-sixth of the surface gravity:
\be
\kappa = 6\frac{\p{}M}{\p\cA}\Big|_{(N,J_n)} = \frac{1}{2r_h} \, ,
\ee
which coincides with the one given in Eq. (\ref{8dAk}). Next, the potential $\psi_h$ and the
quasi-angular momentum $\omega_h$, which are conjugate to $N$ and $J_n$, respectively, are
given by
\bea
&&\psi_h = \frac{\p{}M}{\p{}N}\Big|_{(\cA,J_n)} =
\frac{2nr_h\big(r_h^4 +10n^2r_h^2 -15n^4\big)}{5\big(r_h^2 +n^2\big)} \, , \\
&&\omega_h = \frac{\p{}M}{\p{}J_n}\Big|_{(\cA,N)} = \frac{n}{r_h^2 +n^2} \, .
\eea
Using all the above thermodynamical conjugate pairs, it is easy to check that both differential
and integral mass formulas are completely obeyed
\bea
dM &=& (\kappa/6)d\cA +\omega_h{}dJ_n +\psi_h{}dN \, ,  \label{dmf8d} \\
5M &=& \kappa\cA +6\omega_h{}J_n +\psi_h{}N \, . \label{imf8d}
\eea
Then it is natural to recognize
\be
S = \frac{A_h}{4} = \frac{\pi}{3}\cA = 16\pi^3\big(r^2 +n^2\big)^3 \, , \quad
T = \frac{\kappa}{2\pi} = \frac{1}{4\pi{}r_h} \, ,
\ee
so that the eight-dimensional Taub-NUT solution behaves like a genuine black hole without
violating the beautiful one-quarter area-entropy law. Here we do not require in advance that
the first law should be obeyed in order to obtain the consistent thermodynamical relations,
rather it is just a very natural by-product of the purely algebraic deduction.

\subsection{Extension to the Taub-NUT-AdS$_8$ spacetime}\label{IIIB}

In this subsection, we would like to deal with the Lorentzian Taub-NUT-AdS$_8$ spacetime with
a nonzero cosmological constant. The metric is still given by Eq. (\ref{8dNUT}), but now
\bea
f(r) &=& \frac{1}{5\big(r^2 +n^2\big)^3}\Big[r^6 +5n^2r^4 +15n^4r^2 -5n^6 -10mr \nn \\
 && +g^2\big(5r^8 +28n^2r^6 +70n^4r^4 +140n^6r^2 -35n^8\big)\Big] \, , \nn
\eea
in which $l = 1/g$ is the cosmological scale.

First, one can evaluate the AD mass for this spacetime as:
\be
M = 48\pi^2m \, .
\ee

Next, we want to compute some thermodynamic quantities at the Killing horizon that is
determined by $f(r_h) = 0$. At the horizon, the surface gravity can be obtained via the
standard method as
\be
\kappa = \frac{1}{2}f^{\prime}(r_h) = \frac{1 +7g^2\big(r_h^2 +n^2\big)}{2r_h} \, ,
\label{8dkappa}
\ee
while the horizon area is $A_h = 64\pi^3\cA_h$, with the reduced horizon area still being
denoted as $\cA_h = (r_h^2 +n^2)^3$.

Then we substitute $r_h = \sqrt{\cA_h^{1/3} -n^2}$ into the equation: $[r_h^6 +5n^2r_h^4
+15n^4r_h^2 -5n^6 +g^2(5r_h^8 +28n^2r_h^6 +70n^4r_h^4 +140n^6r_h^2 -35n^8)]^2 = 100m^2r_h^2$
to get an identity:
\bea
m^2 &=& \frac{1}{100\cA_h^{1/3}}\Big[\big(1 +8g^2n^2\big)\big(\cA_h +2n^2\cA_h^{2/3}
 +8n^4\cA_h^{1/3} \nn \\
&&-16n^6\big) +5g^2\cA_h^{4/3}\Big]^2 +\frac{m^2n^2}{\cA_h^{1/3}} \, . \label{8dsqmAdS}
\eea
Supposed that only the secondary hair $J_n = Mn$ is needed to be included as before, then
after inserting $m = M/(48\pi^2)$, $n = N/(48\pi^2)$, $\cA = 48\pi^2\cA_h$ and $g^2 = 8\pi{}
P/21$ into Eq. (\ref{8dsqmAdS}), one can arrive at the following squared-mass formula:
\bea\label{8dsqMAdS}
M^2 &=& \frac{\big(6\pi^2\big)^{1/3}}{50\cA^{1/3}}\bigg\{\Big(1 +\frac{N^2}{756\pi^3}P\Big)
\Big[\cA +\frac{N^2\big(6\pi^2\cA^2\big)^{1/3}}{576\pi^4} \nn \\
&&+\frac{N^4(36\pi\cA)^{1/3}}{165888\pi^7} -\frac{N^6}{15925248\pi^{10}}\Big]^2 \nn \\
&&+\frac{10}{63}\big(36\pi\cA^4\big)^{1/3}P \bigg\}^2
 +\frac{2\big(6\pi^2\big)^{1/3}}{\cA^{1/3}}J_n^2 \, ,
\eea
in which $P$ is the generalized pressure. We point out that the squared-mass formula
(\ref{8dsqMAdS}) consistently reduces to Eq. (\ref{8dsqM}) when the cosmological constant
vanishes.

Similar to the strategy as did in the last subsection, one can view the mass as an implicit
function: $M = M(\cA, N, J_n, P)$, and then differentiating the squared-mass formula
(\ref{8dsqMAdS}) with respect to its variables leads to a new reasonable differential
mass formula:
\be
dM = (\kappa/6)d\cA +\omega_h{}dJ_n +\psi_h{}dN +VdP \, ,  \label{8dFL}
\ee
where
\bea
\kappa &=& 6\frac{\p{}M}{\p{}\cA}\Big|_{(N,J_n,P)}
 = \frac{1 +7g^2\big(r_h^2 +n^2\big)}{2r_h} \, , \nn \\
\omega_h &=& \frac{\p{}M}{\p{}J_n}\Big|_{(\cA,N,P)} = \frac{n}{r_h^2 +n^2} \, , \nn \\
\psi_h &=& \frac{\p{}M}{\p{}N}\Big|_{(\cA,J_n,P)}
 = \frac{2nr_h}{5\big(r_h^2 +n^2\big)} \Big[r_h^4 +10n^2r_h^2 -15n^4 \nn \\
&&\qquad\qquad +4g^2\big(r_h^6 +7n^2r_h^4 +35n^4r_h^2 -35n^6\big)\Big] \, , \nn \\
V &=& \frac{\p{}M}{\p{}P}\Big|_{(\cA,N,J_n)} \nn \\
&=& \frac{64\pi^3r_h(5r_h^8 +28n^2r_h^6 +70n^4r_h^4
 +140n^6r_h^2 -35n^8)}{35\big(r_h^2 +n^2\big)} \, . \nn
\eea
At the same time, one can check that the integral mass formulas
\be
5M = \kappa\cA +6\omega_h{}J_n +\psi_h{}N -2VP \, ,
\ee
is also automatically satisfied.

The consistency of the above thermodynamic relations suggests that one should restore the
well-known Bekenstein-Hawking area-entropy relation $S = A_h/4 = 16\pi^3\cA_h$ and Hawking-Gibbons
temperature $T = \kappa/(2\pi)$, which means that the eight-dimensional Taub-NUT-AdS spacetime
should be regarded as a generic black hole.

It is worth to note that the thermodynamic quantities of the base space of $S^2\times\CP2$
are the same ones as those in the case of $S^2 \times{} S^2 \times{} S^2$ base space, because
the the expression of the radial function $f(r)$ remains unchanged, and we will not repeat
them here.

\section{10-dimensional Taub-NUT spacetime}\label{s4}

Finally, we will turn to consider the 10-dimensional Taub-NUT spacetime and its AdS counterpart.
As shown in Ref. \cite{CQG19-2051} for the 10-dimensional Taub-NUT spacetime, there are three
different choices for the base manifold, namely $S^2 \times{} S^2 \times{} S^2\times{} S^2$,
$S^2\times{} S^2\times\CP2$, and $\CP2\times \CP2$. We will only consider the case in which
the metric possesses a $U(1)$ fibration over $S^2 \times{} S^2 \times{} S^2\times{} S^2$:
\bea
ds_{10}^2 &=& -f(r)\Big(dt +2n\sum_{i=1}^4\cos\theta_i{}d\phi_i\Big)^2 +\frac{dr^2}{f(r)} \nn \\
&&+(r^2 +n^2)\sum_{i=1}^4\big(d\theta_i^2 +\sin^2\theta_i{}d\phi_i^2\big) \, , \label{10dNUT}
\eea
where

\bea
f(r) = \frac{5r^8 +28n^2r^6 +70n^4r^4 +140n^6r^2 -35n^8 -70mr}{35\big(r^2 +n^2\big)^4} \, . \nn
\eea

At the horizon which is defined by the largest root of $f(r_h) = 0$, the horizon area and the
surface gravity can be obtained as
\be
A_h = 256\pi^4\big(r_h^2 +n^2\big)^4 = 256\pi^4\cA_h \, , \quad
\kappa = \frac{1}{2}f^{\prime}(r_h) = \frac{1}{2r_h} \, ,   \label{10dAk}
\ee
where the reduced area is denoted as: $\cA_h = (r_h^2 +n^2)^4$.

The expressions of the AD mass and the NUT charge can be similarly calculated as
\be
M = 256\pi^3m \, , \qquad N = 256\pi^3n \, .
\ee

\subsection{Consistent mass formulas of the 10-dimensional Taub-NUT spacetime}

Adopting the same strategy as did before, we insert $r_h = \sqrt{\cA_h^{1/4} -n^2}$ into
the equation: $(5r_h^8 +28n^2r_h^6 +70n^4r_h^4 +140n^6r_h^2 -35n^8)^2 = 4900m^2r_h^2 $,
and after some computations, we can get an useful identity:
\bea
m^2 &=& \frac{1}{4900\cA_h^{1/4}}\big(5\cA_h +8n^2\cA_h^{3/4} +16n^4\cA_h^{1/2} \nn \\
&&+64n^6\cA_h^{1/4} -128n^8 \big)^2 +\frac{m^2n^2}{\cA_h^{1/4}} \, . \label{10dsqm}
\eea
After substituting $m = M/(256\pi^3)$, $n = N/(256\pi^3)$, $\cA = 256\pi^3\cA_h$ and the
secondary hair $J_n =  Mn$ into Eq. (\ref{10dsqm}), one can obtain the following squared-mass
formula:
\bea
M^2 &=& \frac{\pi^{3/4}}{49\cA^{1/4}}\bigg[\cA +\frac{N^2(\pi\cA)^{3/4}}{10240\pi^6}
 +\frac{N^4\sqrt{\pi\cA}}{83886080\pi^{11}} \nn \\
&&+\frac{N^6(\pi\cA)^{1/4}}{343597383680\pi^{16}}
-\frac{N^8}{2814749767106560\pi^{21}}\bigg]^2 \nn \\
&&+\frac{4\pi^{3/4}}{\cA^{1/4}}J_n^2 \, . \label{10dsqM}
\eea

In the following, the differential and integral mass formulas for the ten-dimensional Taub-NUT
spacetime will be derived under the assumption that the entire set of thermodynamic quantities
is: the mass $M$, the NUT charge $N$, and the secondary hair $J_n = Mn$, which will also be
viewed as an independent variable. Differentiating the squared-mass formula (\ref{10dsqM})
with respect to $\cA$ yields one-eighth of the surface gravity:
\be
\kappa = 8\frac{\p{}M}{\p\cA}\Big|_{(N,J_n)} = \frac{1}{2r_h} \, ,
\ee
which is accordance with the one given in Eq. (\ref{10dAk}). The gravito-magnetic potential
$\psi_h$ and the quasi-angular momentum $\omega_h$, which are conjugate to $N$ and $J_n$,
respectively, can be computed as
\bea
\psi_h &=& \frac{\p{}M}{\p{}N}\Big|_{(\cA, J_n)}
= \frac{8nr_h\big(r_h^6 +7n^2r_h^4 +35n^4r_h^2
 -35n^6\big)}{35\big(r_h^2 +n^2\big)} \, , \qquad  \\
\omega_h &=& \frac{\p{}M}{\p{}J_n}\Big|_{(\cA, N)} = \frac{n}{r_h^2 +n^2} \, .
\eea

One can readily verify that both the differential and integral mass formulas
\bea
dM &=& (\kappa/8)d\cA +\omega_h{}dJ_n +\psi_h{}dN \, ,  \label{dmf10d} \\
7M &=& \kappa\cA +8\omega_h{}J_n +\psi_h{}N \, , \label{imf10d}
\eea
are fully obeyed by using all the thermodynamical conjugate pairs given above. It is natural
to identify
\be
S = \frac{A_h}{4} = \frac{\pi}{4}\cA = 64\pi^4\big(r_h^2 +n^2\big)^4 \, , \quad
T = \frac{\kappa}{2\pi} = \frac{1}{4\pi{}r_h} \, ,
\ee
so that the ten-dimensional Taub-NUT solution acts like a true black hole without violating
the beautiful one-quarter area-entropy relation. Here, we do not require ahead that the first
law be obeyed to achieve consistent thermodynamical connections, rather, it is a very natural
by-product of purely algebraic deduction.

\subsection{Extension to the Taub-NUT-AdS$_{10}$ spacetime}

Finally we would like to tackle with the Lorentzian Taub-NUT-AdS$_{10}$ spacetime with a
nonzero cosmological constant. The metric is still given by Eq. (\ref{10dNUT}), and now
we have
\bea
f(r) &=& \frac{1}{35\big(r^2 +n^2\big)^4}\Big[5r^8 +28n^2r^6 +70n^4r^4 +140n^6r^2 \nn \\
&&-35n^8-70mr +5g^2\big(7r^{10} +45n^2r^8 +126n^4r^6 \nn \\
&&+210n^6r^4 +315n^8r^2 -63n^{10}\big)\Big] \, , \nn
\eea
where $l = 1/g$ is the cosmological scale.

Similar to the low dimensional case, one can compute the AD mass for this spacetime as:
\be
M = 256\pi^3m \, .
\ee

Below, we will evaluate some thermodynamic quantities related to the Killing horizon which
is specified by $f(r_h) = 0$. The surface gravity at the horizon is easily obtained via the
standard method as
\be
\kappa = \frac{1}{2}f^{\prime}(r_h) = \frac{1 +9g^2\big(r_h^2 +n^2\big)}{2r_h} \, ,
\label{10dAdSkappa}
\ee
and the event horizon area still reads $A_h = 256\pi^4\cA_h$, in which the reduced horizon
area is $\cA_h = (r_h^2 +n^2)^4$.

Now it is a position to derive a novel squared-mass formula. Inserting $r_h = \sqrt{\cA_h^{1/4}
-n^2}$ into the equation: $[5r_h^8 +28n^2r_h^6 +70n^4r_h^4 +140n^6r_h^2 -35n^8 +5g^2(7r_h^{10}
+45n^2r_h^8 +126n^4r_h^6 +210n^6r_h^4 +315n^8r_h^2 -63n^{10})]^2 = 4900m^2r_h^2$, and after a
little \newpage\noindent
algebra, we can obtain a useful identity:
\bea
m^2 &=& \frac{1}{4900\cA_h^{1/4}}\Big[\big(1 +10g^2n^2\big)\big(5\cA_h +8n^2\cA_h^{3/4}
 +16n^4\cA_h^{1/2} \nn \\
&&+64n^6\cA_h^{1/4} -128n^8\big) +35g^2\cA_h^{5/4}\Big]^2
 +\frac{m^2n^2}{\cA_h^{1/4}} \, . \label{10dAdSsqm}
\eea
Then after plugging $m = M/(256\pi^3)$, $n = N/(256\pi^3)$, $\cA = 256\pi^3\cA_h$, and $g^2 =
2\pi{}P/9$ into Eq. (\ref{10dAdSsqm}), where $P$ is the generalized pressure, and the secondary
hair: $J_n = Mn$, one can get the following identity:
\bea
M^2 &=& \frac{\pi^{3/4}}{49\cA^{1/4}}\bigg\{\Big(1 +\frac{5N^2}{147456\pi^5}P\Big)
\Big[\cA +\frac{N^2(\pi\cA)^{3/4}}{10240\pi^6} \nn \\
&&+\frac{N^4\sqrt{\pi\cA}}{83886080\pi^{11}}
 +\frac{N^6(\pi\cA)^{1/4}}{343597383680\pi^{16}} \nn \\
&&-\frac{N^8}{2814749767106560\pi^{21}}\Big] +\frac{7}{18\pi}(\pi\cA)^{5/4}P \bigg\}^2 \nn \\
&&+\frac{4\pi^{3/4}}{\cA^{1/4}}J_n^2 \, , \label{10dAdSsqM}
\eea
which is the Christodoulou-Ruffini-like squared-mass formula for the ten-dimensional
Taub-NUT-AdS spacetime. We again point out that this squared-mass formula consistently
reduces to the one obtained in Eq. (\ref{10dsqM}) when the generalized pressure $P$ is
turned off.

Now, as did before, one can regard the mass $M$ as an elementary function: $M = M(\cA, N,
J_n, P)$, and then after differentiating the squared-mass formula (\ref{10dAdSsqM}) with
respect to its variables, one can obtain a reasonable differential mass formula:
\be
dM = (\kappa/8)d\cA +\omega_h{}dJ_n +\psi_h{}dN +VdP \, ,  \label{10dAdSFL}
\ee
where
\bea
\kappa &=& 8\frac{\p{}M}{\p\cA}\Big|_{(N,J_n,P)}
 = \frac{1 +9g^2\big(r_h^2 +n^2\big)}{2r_h} \, , \nn \\
\omega_h &=& \frac{\p{}M}{\p{}J_n}\Big|_{(\cA,N,P)} = \frac{n}{r_h^2 +n^2} \, , \nn \\
\psi_h &=& \frac{\p{}M}{\p{}N}\Big|_{(\cA,J_n,P)} \nn \\
&=& \frac{2nr_h}{35\big(r_h^2 +n^2\big)}\Big[4\big(r_h^6 +7n^2r_h^4
 +35n^4r_h^2 -35n^6\big) \nn \\
&&+5g^2\big(5r_h^8 +36n^2r_h^6 +126n^4r_h^4 +420n^6r_h^2 -315n^8\big)\Big] \, , \nn\\
V &=& \frac{\p{}M}{\p{}P}\Big|_{(\cA,N,J_n)} \nn \\
&=& \frac{256\pi^4r_h}{63\big(r_h^2 +n^2\big)}\big(7r_h^{10} +45n^2r_h^8 +126n^4r_h^6
 +210n^6r_h^4 \nn \\
&&+315n^8r_h^2 -63n^{10}) \, . \nn
\eea
In the meanwhile, one can easily verify that the Bekenstein-Smarr mass formula
\be
7M = \kappa\cA +8\omega_h{}J_n +\psi_h{}N -2VP \, ,\label{10dAdSBS}
\ee
is completely satisfied also.

Comparing our new mass formulas as displayed in Eqs. (\ref{10dAdSFL})-(\ref{10dAdSBS})
with the familiar standard ones, it is strongly suggested that one should make the familiar
identifications $S = A_h/4 = 64\pi^4\cA_h$ and $T = \kappa/(2\pi)$, which restores the famous
Bekenstein-Hawking one-quarter area-entropy relation of the ten-dimensional Taub-NUT-AdS
spacetime in a very pleasing way, so that the solution behaves like a genuine black hole.

Here, we also point out that thermodynamic quantities in the cases of $S^2\times{} S^2\times
\CP2$ and $\CP2\times \CP2$ base space should be the same ones as those in the case of $S^2
\times{} S^2 \times{} S^2 \times{} S^2$ base manifold since the expression of the radial
function $f(r)$ remains unchanged, so we will not present them.

\section{Summary: general ($2k+2$)-dimensional cases}\label{s5}

To summarize, we have established the consistent thermodynamic first law and Bekenstein-Smarr
mass formula for the generic $D = (2k+2)$ Lorentzian Taub-NUT (AdS) spacetimes whose metrics
are compactly written as
\bea
ds_D^2 &=& -f(r)\Big(dt +2n\sum_{i=1}^k\cos\theta_i{}d\phi_i\Big)^2 +\frac{dr^2}{f(r)} \nn \\
&&+(r^2 +n^2)\sum_{i=1}^k\big(d\theta_i^2 +\sin^2\theta_i{}d\phi_i^2\big) \, ,
\eea
with the radial function being
\bea
f(r) &=& \bigg\{\int^r\big[1 +(2k+1)g^2\big(x^2+n^2\big)\big]
\frac{\big(x^2+n^2\big)^k}{x^2}dx    \nn \\
&&-2m\bigg\}\frac{r}{\big(r^2+n^2\big)^k} \, . \nn
\eea

These higher even-dimensional Taub-NUT-AdS spacetimes are shown to be subject to the
traditional forms of the first law and the Bekenstein-Smarr mass formula as follows
\bea
dM &=& TdS +\omega_h{}dJ_n +\psi_h{}dN +VdP \, , \label{FLg}\\
(D-3)M &=& (D-2)(TS +\omega_h{}J_n) +\psi_h{}N -2VP \, , \label{BSg}
\eea
provided that a new secondary hair: $J_n = Mn$ is included just like in the case of their
four-dimensional cousins \cite{PRD100-101501,PRD105-124013}.

The thermodynamical quantities that enter the above differential and integral mass formulas
are given below
\bea
M &=& k(4\pi)^{k-1}m \, , \qquad N = k(4\pi)^{k-1}n \, , \nn \\
J_n &=& k(4\pi)^{k-1}mn \, , \qquad
S = \frac{1}{4}\big[4\pi\big(r_h^2+n^2\big)\big]^k \, , \nn \\
T &=& \frac{f^{\prime}(r_h)}{4\pi}
 = \frac{1 +(2k+1)g^2\big(r_h^2+n^2\big)}{4\pi{}r_h} \, , \nn \\
\omega_h &=& \frac{n}{r_h^2+n^2} \, , \qquad P = \frac{k(2k+1)}{8\pi}g^2 \, , \nn \\
V &=& \frac{(4\pi)^k r_h^2}{r_h^2+n^2}\int^{r_h}\frac{\big(x^2+n^2\big)^{k+1}}{x^2}dx \, , \nn \\
\psi_h &=& -\frac{1+(2k+1)g^2\big(r_h^2+n^2\big)}{2nr_h}\big(r_h^2+n^2\big)^k \nn
\eea
\bea
&&+\frac{(2k-1)r_h^2 -n^2}{2n\big(r_h^2+n^2\big)}
 \int^{r_h}\frac{\big(x^2+n^2\big)^k}{x^2}dx \nn \\
&&+(2k+1)g^2\frac{(2k+1)r_h^2-n^2}{2n\big(r_h^2+n^2\big)}\int^{r_h}
\frac{\big(x^2 +n^2\big)^{k+1}}{x^2}dx \, . \nn
\eea
By the way, the squared mass formulas can be written as:
\bea
&&M^2 = \frac{J_n^2}{4\pi(4S)^{1/k}}
 +\frac{k^2(4\pi)^{2k-1}}{4(4S)^{1/k}}\bigg\{g^2\frac{(4S)^{1+1/k}}{(4\pi)^{k +1}} \nn \\
&&\qquad +\big[1 +2(k+1)g^2n^2\big]r_h\int^{r_h}
 \frac{\big(x^2+n^2\big)^k}{x^2}dx \bigg\}^2 \, ,
\eea
and the following identity must be used to verify that both mass formulas are indeed
fulfilled:
\be
m = \int^{r_h}\big[1 +(2k+1)g^2\big(x^2+n^2\big)\big]
\frac{\big(x^2+n^2\big)^k}{2x^2}dx \, .
\ee
Incidentally, we should point out the four-dimensional NUT-charged case previously discussed
in \cite{PRD100-101501} without the inclusion of the dual mass can be enclosed as a special
case in the above general expressions.

\section{The extra hair $J_n$ as a redundant variable}\label{s6}

In the previous three sections (\ref{s2}-\ref{s4}) which are summarized in Sec. \ref{s5},
by introducing an extra secondary hair: $J_n = Mn$ which has been viewed as a independent
thermodynamic variable just like the 4-dimensional case \cite{PRD100-101501,PRD105-124013},
not only can the traditional thermodynamical first law and Bekenstein-Smarr mass formula
be perfectly extended to the higher even-dimensional NUT-charged cases, but also their
thermodynamical conjugate pairs can be fairly subject to the common Maxwell relations.

However, one might object to our above measure adopted in the last sections (\ref{s2}-\ref{s4})
and doubt that there exists a mathematical inconsistence in our preceding treatments, which
would be a key flaw in that it fails to properly account for the number of independent
parameters appearing in the solutions. In other words, there is a mismatch between the
number of independent solution parameters and that of thermodynamical variables after
introducing an extra secondary hair $J_n$, since it obviously enlarges by one between these
numbers. This can be easily explained by counting the number of the solution parameter space
and that of thermodynamical parameter space as follows. Note that in the usual NUT-less case,
the horizon equation: $f(r_h) = 0$ and the variant $\delta\!{}f(r_h) = 0$ with respect to
its variables imply the Bekenstein-Smarr mass and the first law, respectively, and this
is completely equivalent to deriving both mass formulas from the squared mass formulas and
no mismatch problem arises when both methods are used. Consider now the NUT-charged case, the
equation $f(r_h) = 0$ means that its roots can be written as: $r_h = r_h(m,n,q,g)$, which in
turn can be expressed as an entropy function: $S = S(M,N,Q,P)$. According to the traditional
view if no extra hair is included, then the entropy expression should be converted into a mass
function: $M = M(S,N,Q,P)$, and nothing more is added by hand. However, differently from the
usual practice, we have advocated to include a new secondary hair $J_n$ into the squared mass
formula in the above manipulations, which results in a function relation: $M = M(S,J_n,N,Q,P)$
by enlarging one more parameter into the thermodynamical state space. This apparently leads
to a conflict about the independent freedom of degree since there are only two free parameters
among three thermodynamical variables: ($M, N$ and $J_n $), due to the equality: $J_n = Mn =
(4\pi)^{1-k}MN/k$.

To resolve the contradiction about the mismatch between the number of independent solution
parameters and that of the thermodynamical variables, below we will provide a simple recipe
to deal with this conflict by waiving the import of the secondary hair $J_n$, so that our
preceding treatments would be viewed as a simpler intermediate step towards deriving the
following reduced version of the mass formulas.

Consider $J_n = Mn$ as a redundant variable, that is to say, $J_n$ is not a independent variable
so that we abandon to include $J_n$ as a new hair. Previously, the impact of this constraint on
the thermodynamical relations had already been addressed in the four-dimensional NUT-charged cases
in our papers \cite{PRD100-101501,PRD105-124013}, but ignored in the last sections for their
higher-dimensional versions. Here we shall discuss this issue and derive the corresponding
reduced mass formulas of the general ($2k+2$)-dimensional cases.

Now using $J_n = (4\pi)^{1-k}MN/k$, we can obtain the differentiation: $k(4\pi)^{k-1}dJ_n =
MdN +NdM$ by taking into account $N = k(4\pi)^{k-1}J_n/M$. With the help of these expressions,
we can further eliminate $J_n$ and $dJ_n$ from the differential and integral mass formulas.
Thus, the first law (\ref{FLg}) and Bekenstein-Smarr mass formula (\ref{BSg}) boil down to
their nonstandard forms as follows:
\bea
\Big[1 -\frac{N\omega_h}{k(4\pi)^{k-1}} \Big]dM
&=& TdS +\bar{\psi_h} dN +VdP \, ,  \label{rmf1} \\
 (D-3)\Big[1 -\frac{N\omega_h}{k(4\pi)^{k-1}} \Big]M
&=& (D-2)TS +\bar{\psi}_h N -2VP \label{rmf2} \, , \qquad
\eea
where $\bar{\psi}_h = \psi_h +(4\pi)^{1-k}M\omega_h/k$.

It is easy to see that all of the thermodynamic quantities in the reduced mass formulas
(\ref{rmf1}-\ref{rmf2}) cannot constitute the ordinary canonical conjugate pairs and do
not obey the conventional Maxwell relations due to the presence of a pre-factor $\big[1
-(4\pi)^{1-k}N\omega_h/k\big]$ in front of $dM$ and $M$. A similar situations previously
appeared in the four-dimensional superentropic Kerr-Newman-AdS, ultraspinning Kerr-Sen-AdS and
ultraspinning dyonic Kerr-Sen-AdS black holes \cite{PRD103-044014,PRD101-024057,PRD102-044007,
JHEP0114127,PRD89-084007,PRL115-031101,JHEP0615096}, where the chirality condition: $J = Ml$
($l=1/g$ is the cosmological scales) reduces one of the numbers of independent thermodynamical
parameters of their corresponding usual black holes after taking the $a\to{}l$ limit, so that
the standard forms of usual thermodynamics are reduced to the nonstandard relations.

Finally, we can also note that the squared mass formula is recast into
\bea
&&\Big[k^2(4S)^{1/k} -(4\pi)^{1-2k}N^2\Big]M^2
 =\frac{k^4(4\pi)^{2k-1}}{4}\bigg\{g^2\frac{(4S)^{1+1/k}}{(4\pi)^{k +1}} \nn
\eea
\bea
&&\quad +\big[1 +2(k+1)g^2n^2\big]r_h\int^{r_h}
 \frac{\big(x^2+n^2\big)^k}{x^2}dx \bigg\}^2 \, .\qquad
\eea

\section{Conclusions and outlooks}\label{s7}

In our previous work \cite{PRD100-101501,PRD105-124013}, we have suggested from the thermodynamical
perspective that the NUT charge behave like a thermodynamical multi-hair in the mass formulas of
the four-dimensional NUT-charged spacetimes, of which a great advantage is that not only both the
integral and differential mass formulas inherit the conventional forms in an elegant way, but also
the thermodynamical quantities constitute the usual relations of common conjugate pairs. What is
more, both the famous Bekenstein-Hawking one-quarter of area-entropy relation $S = A_h/4$ and the
Hawking-Gibbons temperature formula $T =\kappa/(2\pi)$ can be naturally applied to all NUT-charged
spacetimes. These are the most striking differences from other relevant attempts of the mainstream
community \cite{PRD100-064055,JHEP0719119,CQG36-194001,PLB798-134972,JHEP0520084,PRD100-104016,
PLB832-137264,PLB802-135270,IJMPD31-2250021,JHEP0821152,PRD101-124011,PRD105-124034,PRD103-024052,
JHEP0321039,PRD106-024022,EPJP130-124,2112.00780,2208.05494}. On the other hand, the novelty of
our proposal is that it not only aims to copy with thermodynamical aspect, but also takes account
of other properties, such as the explanation of the gyromagnetic ratio and the quantization
condition for a gravitational monopole. In particular, without considering the secondary hair
$J_n = Mn$ as a independent charge, the universal rule of the area (entropy) products cannot be
applied to the NUT-charged spacetimes \cite{PLB807-135521}.

In this paper, we have adopted the same strategy and successfully achieved the consistent first
law and Bekenstein-Smarr mass formula for the six-, eight-, and ten-dimensional Lorentzian Taub-NUT
(AdS) spacetimes. Up to date, our work is the only one to deal with thermodynamics of higher
even-dimensional Lorentzian Taub-NUT (AdS) spacetimes. Similar to the cases of the four-dimensional
Lorentzian Taub-NUT (AdS) solutions, as did in our previous works \cite{PRD100-101501,PRD105-124013},
we also import only one secondary hair: $J_n = Mn$ here. A key rudiment of this work is to deduce
a reasonable Christodoulou-Ruffini-like squared-mass formula for each dimension, which represents
a hyper-surface in one more high-dimensional thermodynamical state space. From this squared-mass
formula, the thermodynamical first law and Bekenstein-Smarr mass formula can be derived via simple
differentiations with respect to its thermodynamic variables, and the resultant thermodynamical
conjugate pairs meet their standard forms of the differential and integral mass formulas. After
collecting all main results in a compact fashion for the generic ($2k+2$)-dimensional Lorentzian
NUT-charged spacetimes, we then have dealt with the case when the secondary hair $J_n = Mn$ is not
viewed as a independent variable so as to resolve a potential mathematical inconsistence behind
in our preceding prescription. We should mention that all the results obtained in this paper
resembles to the cases of the four-dimensional Lorentzian Taub-NUT (AdS) spacetime, however there
is an exception in that the notion of a dual (magnetic) mass in higher dimensions is currently
unclear to be defined. Once an appropriate definition for it is proposed, our present work might
be modified accordingly via the further inclusion of it.

Our study in this paper demonstrated that our idea ``The NUT charge is a thermodynamical multi-hair"
has a universal applicability, and our method is effective and systematical. A natural question is:
whether it is applicable to deal with the charged versions of the higher even-dimensional Taub-NUT
spacetimes \cite{CQG23-2849,PRD73-124039}. A preliminary research shows that only including one
secondary hair $J_n = Mn$ is not sufficient to resolve the consistency of the first law and integral
mass formula, so at least one more charge should be added into them. For more details, please see our
recent work \cite{2306.00062} about the electrically charged extension of the present paper. Another
related issue is: whether the present work can be extended to treat thermodynamics of the higher
even-dimensional multi-NUTty spacetimes \cite{PLB593-218,CQG21-2937,PLB634-448}, since the solutions
studied in this paper can be viewed as a special equal-NUT case of these more general spacetimes with
multi-NUT parameters. The answer to this question is affirmative, please see the Appendix \ref{app}
for the brief results in the cases without a cosmological constant. We hope to report the details of
the related work soon.

\acknowledgments

We are greatly indebted to the anonymous referees for helpful comments to improve the presentations
of this work. This work is supported by the National Natural Science Foundation of China (NSFC)
under Grant No. 12205243, No. 11675130, by the Sichuan Science and Technology Program under Grant
No. 2023NSFSC1347, and by the Doctoral Research Initiation Project of China West Normal University
under Grant No. 21E028.

\appendix

\section{Consistent thermodynamics of the
($2k+2$)-dimensional multi-NUTty spacetimes}\label{app}

In this Appendix, we will briefly give the main results of the consistent thermodynamics of
the $D = (2k+2)$-dimensional Lorentzian multi-NUTty spacetimes without a cosmological constant.
Using the base spaces $\prod_{i=1}^k\bigotimes{}S^2$, the line elements of these multi-NUTty
spacetimes are written as \cite{PLB634-448}:
\bea
ds_D^2 &=& -f(r)\Big(dt +2\sum_{i=1}^kn_i\cos\theta_i{}d\phi_i\Big)^2 +\frac{dr^2}{f(r)} \nn \\
&& +\sum_{i=1}^k(r^2 +n_i^2)\big(d\theta_i^2 +\sin^2\theta_i{}d\phi_i^2\big) \, ,
\eea
with the radial function being
\bea
f(r) = \frac{r}{\prod_{i=1}^k\big(r^2+n_i^2\big)}\bigg\{\int^r
\frac{\prod_{i=1}^k\big(x^2+n_i^2\big)}{x^2}dx -2m\bigg\} \, . \nn
\eea

These multi-NUTty spacetimes obey the usual forms of the first law and the Bekenstein-Smarr
mass formula as follows
\bea
dM &=& TdS +\sum_{i=1}^k(\omega_i{}dJ_i +\psi_i{}dN_i) \, , \\
(2k-1)M &=& 2k{}TS +\sum_{i=1}^k(2k\omega_i{}J_i +\psi_i{}N_i) \, ,
\eea
provided that we introduce $k$ new secondary hairs: $J_i = Mn_i$.

The thermodynamical quantities that appear in the above differential and integral mass formulas
are
\bea
M &=& k(4\pi)^{k-1}m \, , \quad N_i = k(4\pi)^{k-1}n_i \, , \quad
 J_i = k(4\pi)^{k-1}mn_i \, , \nn \\
S &=& \frac{(4\pi)^k}{4}\prod_{i=1}^k\big(r_h^2+n_i^2\big) \, , \quad
T = \frac{1}{4\pi{}r_h} \, , \quad
\omega_i = \frac{n_i}{r_h^2+n_i^2} \, , \nn
\eea
\bea
\psi_i &=& \frac{n_i}{k}\bigg(\sum_{p=1}^k\frac{r_h^2}{r_h^2+n_p^2}\bigg)
 \int^{r_h}\frac{\prod_{j=1}^k\big(x^2+n_j^2\big)}{x^2\big(x^2+n_i^2\big)}dx \nn \\
&& -\frac{n_i}{k\big(r_h^2+n_i^2\big)}\int^{r_h}\prod_{j=1}^k\big(x^2+n_j^2\big)
 \sum_{p=1}^k\frac{1}{x^2+n_p^2}dx \, , \nn
\eea
where $r_h$ is the largest root of the horizon equation: $f(r_h) = 0$. Incidentally, we would
like to emphasize that throughout this article, all the integration constants in the integral
expressions are set to zero.

\end{document}